\documentclass{ws-procs975x65}

\begin{document}

\title{Sensitivity to sterile neutrino mixings and
the discovery channel at a neutrino factory}

\author{Osamu Yasuda$^*$}

\address{Department of Physics, Tokyo Metropolitan University,\\
Minami-Osawa, Hachioji, Tokyo 192-0397, Japan\\
$^*$E-mail: yasuda\_at\_phys.metro-u.ac.jp}

\begin{abstract}
Sensitivity of a neutrino factory to
various mixing angles in a scheme with one sterile neutrino
is studied using $\nu_e\to\nu_\mu$, $\nu_\mu\to\nu_\mu$,
$\nu_e\to\nu_\tau$ and $\nu_\mu\to\nu_\tau$.
While the ``discovery-channel'' $\nu_\mu\to\nu_\tau$ is
neither useful in the standard three flavor scheme nor very powerful
in the sensitivity study of sterile neutrino mixings,
this channel is important to check unitarity and to probe the new CP phase
in the scheme beyond the standard neutrino mixing framework.
\end{abstract}


\bodymatter

\section{Introduction}
It is known that the deficit of the solar and atmospheric neutrinos
are due to neutrino oscillations among three flavors of neutrinos,
and these observations offer evidence of neutrino masses and
mixings.\cite{Amsler:2008zz}
The standard three flavor framework of neutrino oscillations
are described by six oscillation parameters:
three mixing angles $\theta_{12}$, $\theta_{13}$, $\theta_{23}$,
the two independent mass squared differences
$\Delta m^2_{21}$, $\Delta m^2_{31}$,
and one CP violating phase $\delta$,
where $\Delta m^2_{jk}\equiv m^2_j-m^2_k$ and
$m_j$ stands for the mass of the neutrino
mass eigenstate.
From the solar neutrino data, we have
$(\sin^22\theta_{12}, \Delta m^2_{21}) \simeq (0.86,
8.0\times 10^{-3}\mbox{\rm eV}^2)$, and from
the atmospheric neutrino data we have
$(\sin^22\theta_{23}, \Delta m^2_{21}) \simeq (1.0,
2.4\times 10^{-3}\mbox{\rm eV}^2)$.
On the other hand, only the upper bound on $\theta_{13}$
is known ($\sin^22\theta_{13} \leq 0.19$),\footnote{
In Refs.~\refcite{Fogli:2008jx},
\refcite{Fogli:2008cx},
\refcite{Ge:2008sj},
\refcite{Fogli:2009ce},
\refcite{GonzalezGarcia:2010er}, a
global analysis of the neutrino oscillation data has been performed, in
which a non-vanishing best-fit value for $\theta_{13}$ is found. This result,
however, is compatible with $\theta_{13} = 0$ at less than $2 \sigma$,
and it is not yet statistically significant enough to be taken
seriously.}
and no information on $\delta$ is known at present.

To determine $\theta_{13}$ and $\delta$, various neutrino long
baseline experiments have been proposed\cite{Bandyopadhyay:2007kx},
and the ongoing and proposed future neutrino long baseline experiments
with an intense beam include conventional super (neutrino) beam
experiments such as T2K,\cite{Itow:2001ee}
NO$\nu$A,\cite{Ayres:2004js} LBNE,\cite{Maricic:2010zza}
T2KK,\cite{Ishitsuka:2005qi,Hagiwara:2005pe} the $\beta$ beam
proposal,\cite{Zucchelli:2002sa} which uses a $\nu_e$ ($\bar{\nu}_e$)
beam from $\beta$-decays of radioactive isotopes, and the neutrino
factory proposal,\cite{Geer:1997iz} in which $\bar{\nu}_e$ and
$\nu_\mu$ ($\nu_e$ and $\bar{\nu}_\mu$) are produced from decays of
$\mu^-$ ($\mu^+$).  As in the case of the B
factories\cite{belle:0000,babar:0000}, precise measurements in these
experiments allow us not only to determine the oscillation parameters
precisely but also to probe new physics by looking for deviation from
the standard three flavor scheme.  In particular, test of unitarity is
one of the important subjects in neutrino oscillations, and tau
detection is crucial for that purpose.  Among the proposals for future
long baseline experiments, the neutrino factory facility produces a
neutrino beam of the highest neutrino energy, and it is advantageous
to detect $\nu_\tau$, because of the large cross section at high
energy.

New physics which has been discussed in the context of neutrino
oscillation includes sterile neutrinos,\cite{giunti}
the non-standard interactions
during neutrino
propagation\cite{Wolfenstein:1977ue,Guzzo:1991hi,Roulet:1991sm}, the
non-standard interactions at production and
detection\cite{Grossman:1995wx}, violation of unitarity due to heavy
particles\cite{Antusch:2006vwa}, etc.  These scenarios, except the
non-standard interactions during neutrino propagation, offer
interesting possibilities for violation of three flavor unitarity.
Among these possibilities,
phenomenological bound of unitarity violation
is typically of ${\cal O}$(1\%) in the case of
non-standard interactions at production and detection, and
it is of ${\cal O}$(0.1\%) in the case of unitarity violation
due to heavy particles.\cite{Antusch:2008tz}
On the other hand, the bound in the case of
sterile neutrinos is of
${\cal O}$(10\%) which comes mainly from the
constraints of the atmospheric neutrino data,\cite{Donini:2007yf}
so scenarios with sterile neutrinos seem to be
phenomenologically more promising to look for than other
possibilities of unitarity violation.

In this talk I will discuss phenomenology
of schemes with sterile neutrinos at a neutrino factory.
Schemes with sterile neutrinos have attracted a lot of attention
since the LSND group announced the anomaly which suggest neutrino
oscillations with mass squared difference of
${\cal O}$(1eV$^2$).\cite{Athanassopoulos:1996jb,Athanassopoulos:1997pv,Aguilar:2001ty}.
The reason that we need one extra neutrino to account for LSND
is because
the standard three flavor scheme has only two independent
mass squared differences, i.e.,
$\Delta m^2_{21}=\Delta m^2_\odot\simeq 8\times 10^{-5}$eV$^2$
for the solar neutrino oscillation,
and $|\Delta m^2_{31}|=\Delta m^2_{\text{atm}}\simeq 2.4\times 10^{-3}$eV$^2$
for the atmospheric neutrino oscillation,
and it does not have room for the mass squared difference of ${\cal O}$(1eV$^2$).
And the reason that the extra state has to be sterile neutrino,
which is singlet with respect to the gauge group of the Standard
Model, is because the number of weakly interacting light
neutrinos has to be three from the LEP data.\cite{Amsler:2008zz}
The LSND anomaly has been tested by the MiniBooNE experiment,
and it gave a negative result for neutrino oscillations with
mass squared difference of ${\cal O}$(1eV$^2$).\cite{AguilarArevalo:2007it}
While the MiniBooNE data disfavor the region suggested by LSND,
Ref.~\refcite{Karagiorgi:2009nb} gave the allowed region
from the combined analysis of the LSND and MiniBooNE data,
and it is not so clear whether the MiniBooNE data alone
are significant enough to exclude the LSND region.
On the other hand,
even if the Miniboone data are taken as negative evidence against the
LSND region, there still remains a possibility for sterile neutrino
scenarios whose mixing angles are small enough to satisfy the
constraints of Miniboone and the other negative results.
The effect of these scenarios could reveal
as violation of three flavor unitarity in the future neutrino
experiments.  So in this talk I will discuss sterile neutrino schemes
as one of phenomenologically viable possibilities for unitarity violation,
regardless of whether the LSND anomaly is excluded by the
MiniBooNE data or not.

It has been known that sterile neutrino schemes may
have cosmological problems (see,
e.g., Ref.~\refcite{Smirnov:2006bu}).  However, these
cosmological discussions depend on models and assumptions, and
I will not discuss cosmological constraints in this talk.
Also it has been pointed out\footnote{
I would like to thank J.~E.~Kim for calling my attention to
the possibility of the absorption effect due to the
transition magnetic moments of neutrinos.}
that some sterile neutrino models\cite{Kim:1978xk}
have absorption effects even for neutrino energy below 1TeV,
but I will not take such effects into consideration for simplicity.

\section{Schemes with sterile neutrinos}
\begin{figure}
\begin{center}
\psfig{file=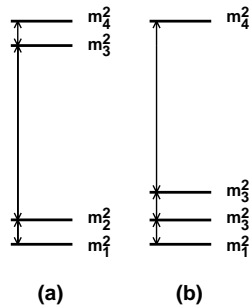,width=0.3\textwidth}
\end{center}
\vspace{-5mm}
\caption{The two classes of four--neutrino mass spectra, (a): (2+2) and
(b): (3+1).}
\label{aba:fig1}
\end{figure}

For simplicity I will discuss schemes with four neutrinos,
although phenomenology of the
schemes with two\cite{Sorel:2003hf} or three\cite{Maltoni:2007zf}
sterile neutrinos have also been
discussed.
Denoting sterile neutrinos as $\nu_s$,
we have the following mixing between the flavor eigen
states $\nu_\alpha~(\alpha=e,\mu,\tau)$ and the
mass eigenstates $\nu_j~(j=1,\cdots,4)$:
\begin{eqnarray}
&{\ }&\left( \begin{array}{c} \nu_e  \\ \nu_{\mu} \\ 
\nu_{\tau}\\\nu_s \end{array} \right)
=\left(
\begin{array}{cccc}
U_{e1} & U_{e2} &  U_{e3} &  U_{e4}\\
U_{\mu 1} & U_{\mu 2} & U_{\mu 3} & U_{\mu 4}\\
U_{\tau 1} & U_{\tau 2} & U_{\tau 3} & U_{\tau 4}\\
U_{s1} & U_{s2} &  U_{s3} &  U_{s4}
\end{array}\right)
\left( \begin{array}{c} \nu_1  \\ \nu_2 \\ 
\nu_3\\\nu_4 \end{array} \right).\nonumber
\end{eqnarray}
There are two kind of schemes with four neutrinos,
depending on how the mass eigenstates are separated by
the largest mass squared difference.
One is the (2+2)-scheme in which two mass eigenstates
are separated by other two, and the other one is
the (3+1)-scheme in which one mass eigenstate is
separated by other three (cf. Fig.\ref{aba:fig1}).

\vglue -0.7cm

\vglue -0.5cm
\subsection{(2+2)-schemes}
In this scheme the fraction of sterile neutrino contributions to solar
and atmospheric oscillations is given by $|U_{s1}|^2+|U_{s2}|^2$ and
$|U_{s3}|^2+|U_{s4}|^2$, respectively, where the mass squared
differences $\Delta m^2_{21}$ and $|\Delta m^2_{43}|$ are assumed to
be those of the solar and atmospheric oscillations.  The experimental
results show that mixing among active neutrinos give dominant
contributions to both the solar and atmospheric oscillations (see,
e.g., Ref.~\refcite{Maltoni:2004ei}).  In particular, in Fig.~19 of
Ref.~\refcite{Maltoni:2004ei} we can see that at the 99\% level
$|U_{s1}|^2+|U_{s2}|^2 \le 0.25$ and $|U_{s3}|^2+|U_{s4}|^2 \le 0.25$,
from the solar and atmospheric oscillations, respectively,
and this contradicts the unitarity condition $\sum_{j=1}^4|U_{sj}|^2= 1$.
In fact the (2+2)-schemes are excluded at 5.1$\sigma$ CL
\cite{Maltoni:2004ei}.  This conclusion is independent of whether we
take the LSND data into consideration or not and I will not consider
(2+2)-schemes in this talk.

\subsection{(3+1)-schemes}
\begin{figure}
\vspace{-3mm}
\hglue -1.0cm
\psfig{file=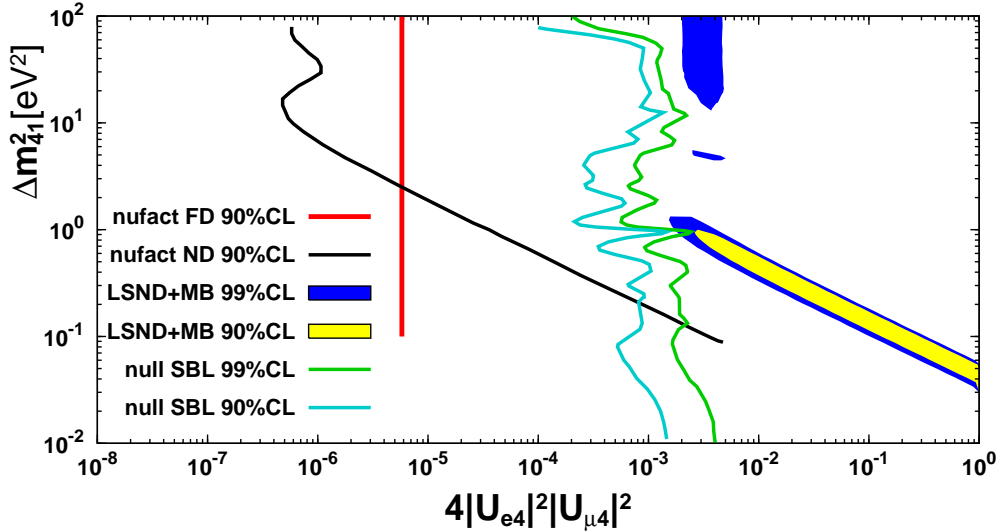,width=1.1\textwidth}
\vglue -0.3cm
\caption{Sensitivity to $4|U_{e4}|^2|U_{\mu4}|^2$
in the (3+1)-scheme
of a 20 GeV neutrino factory with
Far Detectors\cite{Donini:2008wz} (FD)
or with Near Detectors\cite{Donini:2001xy} (ND).
Also shown are the allowed region from the combined analysis of the LSND and
MiniBooNE data\cite{Karagiorgi:2009nb} as well as
the excluded region by all the negative data of
short baseline experiments and atmospheric neutrino
observations.\cite{Maltoni:2004ei,Karagiorgi:2009nb}
}
\label{aba:fig2}
\end{figure}

Phenomenology of the (3+1)-scheme is almost the same as that of
the standard three flavor scenario, as far as the solar
and atmospheric oscillations are concerned.
On the other hand, this scheme has tension
between the LSND data and other negative results of the short baseline
experiments.  Among others, the CDHSW\cite{Dydak:1983zq} and
Bugey\cite{Declais:1994su} experiments
give the bound on $1-P(\nu_\mu\rightarrow\nu_\mu)$ and
$1-P(\bar{\nu}_e\rightarrow\bar{\nu}_e)$, respectively,
and in order for the LSND data to be affirmative,
the following relation has to be satisfied\cite{Okada:1996kw,Bilenky:1996rw}:
\begin{eqnarray}
\sin^22\theta_{\mbox{\rm\tiny LSND}}(\Delta m^2)
<\frac{1}{4}\,\sin^22\theta_{\mbox{\rm\scriptsize Bugey}}(\Delta m^2)
\cdot
\sin^22\theta_{\mbox{\rm\tiny CDHSW}}(\Delta m^2),
\label{relation31}
\end{eqnarray}
where $\theta_{\mbox{\rm\tiny LSND}}(\Delta m^2)$,
$\theta_{\mbox{\rm\tiny CDHSW}}(\Delta m^2)$,
$\theta_{\mbox{\rm\scriptsize Bugey}}(\Delta m^2)$ are the value of
the effective two-flavor mixing angle as a function of the mass
squared difference $\Delta m^2$ in the allowed region for LSND
($\bar{\nu}_\mu\rightarrow\bar{\nu}_e$), the CDHSW experiment
 ($\nu_\mu\rightarrow\nu_\mu$), and the Bugey
experiment 
($\bar{\nu}_e\rightarrow\bar{\nu}_e$), respectively.
The (3+1)-scheme to account for LSND in terms of neutrino oscillations
is disfavored because
eq.~(\ref{relation31}) is not satisfied for any value of $\Delta m^2$.
This argument has been shown quantitatively by
Ref.~\refcite{Maltoni:2004ei} including
the atmospheric neutrino data and other negative results.
In Fig.\ref{aba:fig2} the right hand side of the
lines denoted as ``null SBL 90\% (99\%)''
is the excluded region at 90\% (99\%) CL by
the atmospheric neutrino data and all the negative results
of short baseline experiments, whereas
the allowed region by the combined analysis
of the LSND and MiniBooNE data at 90\% (99\%) CL is also shown.

In the following discussions I will assume the mass pattern depicted in
Fig.\ref{aba:fig1}(b) because the inverted (3+1)-scheme is disfavored by
cosmology, and I will also assume for simplicity that the largest mass
squared difference $\Delta m^2_{41}$ is larger than ${\cal
O}$(0.1eV$^2$), so that I can average over rapid oscillations due to
$\Delta m^2_{41}$ in the long baseline experiments as well as in the
atmospheric neutrino observations.

\section{Sensitivity of a neutrino factory to the sterile neutrino mixings}

\subsection{Neutrino factories}
Unlike conventional long baseline neutrino experiments, neutrino
factories use muon decays $ \mu^+ \rightarrow e^+ \, \nu_e \, \bar
\nu_\mu$ and $\mu^- \rightarrow e^- \, \bar \nu_e \, \nu_\mu $ to
produce neutrinos.  In the setup suggested in the Physics
Report\cite{Bandyopadhyay:2007kx} of {\em
International Scoping Study for a future Neutrino Factory and
Super-Beam facility}, muons of both polarities are
accelerated up to $E_\mu = 20$ GeV and injected into one storage ring
with a geometry that allows to aim at two far detectors, the first
located at 4000 km and the second at 7500 km from the source.  The
reason to put far detectors at two locations is to resolve so-called
parameter
degeneracy.\cite{BurguetCastell:2001ez,Minakata:2001qm,Fogli:1996pv,Barger:2001yr}
The useful channels at neutrino factories are the following:

$\bullet$ $\nu_e\to\nu_\mu$ and $\bar\nu_e\to\bar\nu_\mu$: the golden channel

$\bullet$ $\nu_\mu\to\nu_\mu$ and $\bar\nu_\mu\to\bar\nu_\mu$: the disappearance channel

$\bullet$ $\nu_e\to\nu_\tau$ and $\bar\nu_e\to\bar\nu_\tau$: the silver channel

$\bullet$ $\nu_\mu\to\nu_\tau$ and $\bar\nu_\mu\to\bar\nu_\tau$: the discovery
channel\footnote{It has been known\cite{donini-ids4}
that this channel is
not useful in the standard three flavor framework.
On the other hand, once one starts studying
physics beyond the standard three flavor scenario,
this channel becomes very important.
This is the reason why it is called the discovery
channel.\cite{Donini:2008wz}}

\smallskip
\noindent
At neutrino factories, electrons and positrons produced out of $\nu_e$
and $\bar \nu_e$ create electromagnetic showers, which make it
difficult to identify their charges.  On the other hand,
charge identification is much easier for $\mu$ detection,
so the golden channel
$\nu_e\to\nu_\mu$ is used unlike the conventional
long baseline neutrino experiments which use
$\nu_\mu\to\nu_e$.  The golden channel
turns out to be powerful because of very low backgrounds.
The disappearance channel is also useful
because of a lot of statistics.
The golden and disappearance channels are observed by
looking for muons with magnetized iron
calorimeters.\cite{CerveraVillanueva:2008zz}
On the other hand, the silver and discovery channels are
observed by looking for $\tau$'s with emulsion
cloud chambers (nonmagnetized\cite{Donini:2002rm,Autiero:2003fu}
or magnetized\cite{Abe:2007bi}),
and the statistics of the silver channel is limited.
The silver channel is useful to resolve
parameter degeneracy.
Combination of the golden, disappearance and
discovery channels
is expected to enable us to check unitarity.

\subsection{Sensitivity of a neutrino factory with far detectors\cite{Donini:2008wz}}

Ref.~\refcite{Donini:2008wz} studied sensitivity of
a neutrino factory with far detectors
to sterile neutrino mixings.
The setup is the following:\footnote{
In Ref.~\refcite{Donini:2008wz} an analysis was performed also
for the case of muon energy
50GeV and the baseline lengths $L$=3000km and $L$=7500km,
and it was shown that sensitivity with $\tau$ detectors increases
for 50GeV because of higher statistics.
In this talk, however, I will only mention the results
for the neutrino factory with muon energy 20GeV for simplicity.}
the muon energy is 20GeV,
the number of useful muons is
$5\times10^{20}$ $\mu^-$'s {\it and} $\mu^+$'s per year,
the measurements are supposed to continue
for 4 years, the baseline lengths are $L$=4000km and $L$=7500km,
the volume of each magnetized iron
calorimeter at the two distances is 50kton,
that of each magnetized emulsion cloud chamber
at the two distances is 4kton,
and the statistical as well as systematic errors and
the backgrounds are taken into account.

\begin{figure}[t]
\begin{center}
\psfig{file=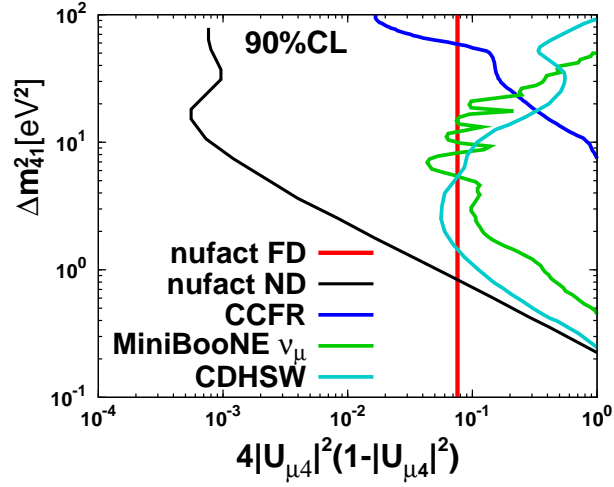,width=0.65\textwidth}
\end{center}
\vglue -0.3cm
\caption{Sensitivity to $4|U_{\mu4}|^2(1-|U_{\mu4}|^2)$
in the (3+1)-scheme
of a 20 GeV neutrino factory with
Far Detectors\cite{Donini:2008wz}
or with Near Detectors\cite{Donini:2001xy}.
The excluded regions by CDHSW\cite{Dydak:1983zq},
by CCFR\cite{Stockdale:1984cg} and by the MiniBooNE $\nu_\mu$
data\cite{Karagiorgi:2009nb} are also shown.
}
\label{aba:fig3}
\end{figure}

\begin{figure}[t]
\begin{center}
\psfig{file=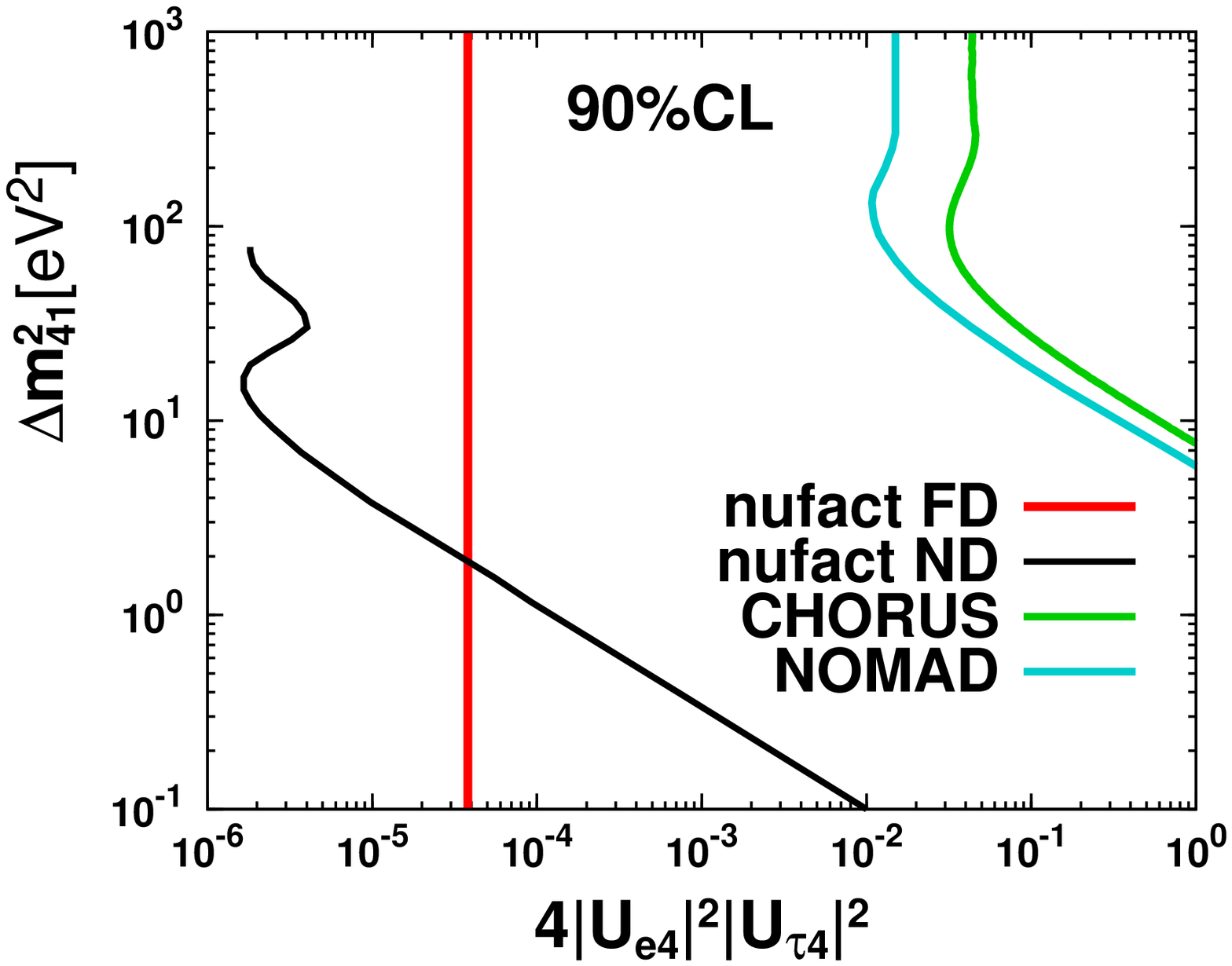,width=0.65\textwidth}
\end{center}
\vglue -3mm
\caption{Sensitivity to $4|U_{e4}|^2|U_{\tau4}|^2$
in the (3+1)-scheme
of a 20 GeV neutrino factory with
Far Detectors\cite{Donini:2008wz}
or with Near Detectors\cite{Donini:2001xy}.
The excluded regions by NOMAD\cite{Astier:2001yj}
and by CHORUS\cite{Eskut:2007rn} are also shown.
}
\label{aba:fig4}
\end{figure}

\begin{figure}[t]
\begin{center}
\psfig{file=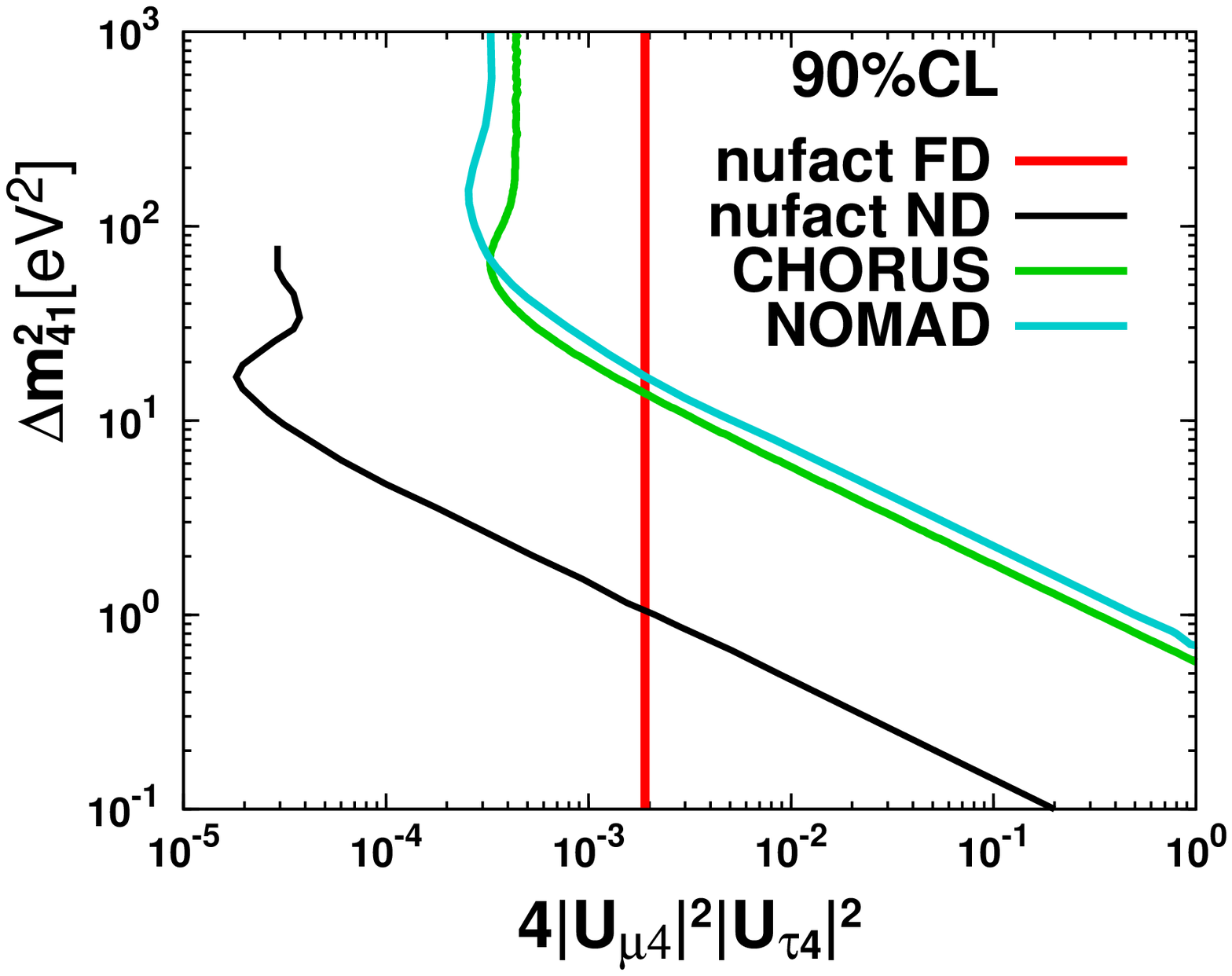,width=0.65\textwidth}
\end{center}
\vglue -3mm
\caption{Sensitivity to $4|U_{\mu4}|^2|U_{\tau4}|^2$
in the (3+1)-scheme
of a 20 GeV neutrino factory with
Far Detectors\cite{Donini:2008wz}
or with Near Detectors\cite{Donini:2001xy}.
The excluded regions by NOMAD\cite{Astier:2001yj}
and by CHORUS\cite{Eskut:2007rn} are also shown.
}
\label{aba:fig5}
\end{figure}

At long baseline lengths such as $L$=7500km,
matter effects become important.
The oscillation probability in constant-density matter
can be obtained by the formalism of
Kimura, Takamura and
Yokomakura\cite{Kimura:2002hb,Kimura:2002wd}.\footnote{
Another proof of the KTY formalism was given in
Refs.~\refcite{Xing:2005gk},\refcite{Yasuda:2007jp}
and it was extended to four neutrino schemes in
Refs.~\refcite{Zhang:2006yq},\refcite{Yasuda:2007jp}.
Analytic forms of the oscillation probability
in the (3+1)-scheme were also given in Ref.~\refcite{Dighe:2007uf}.}
The oscillation probability in matter can be written as
\begin{eqnarray}
\vspace{-5mm}
\hspace{-5mm}
P(\nu_\alpha\rightarrow\nu_\beta)
&=&\delta_{\alpha\beta}
-4\sum_{j<k}\mbox{\rm Re}(\tilde{X}^{\beta\alpha}_j
\tilde{X}^{\beta\alpha\ast}_k)
\sin^2({\Delta \tilde{E}_{jk}L / 2})
\nonumber\\
&{\ }&-2\sum_{j<k}\mbox{\rm Im}(\tilde{X}^{\beta\alpha}_j
\tilde{X}^{\beta\alpha\ast}_k)
\sin(\Delta \tilde{E}_{jk}L),
\nonumber
\end{eqnarray}
\vglue -2mm
\noindent
where
$\tilde{X}^{\alpha\beta}_j\equiv\tilde{U}_{\alpha j}\tilde{U}^\ast_{\beta j}$,
$\Delta \tilde{E}_{jk}\equiv \tilde{E}_j-\tilde{E}_k$,
$\tilde{E}_j$ and $\tilde{U}_{\alpha j}$ are
the energy eigenvalue and the neutrino mixing matrix element in matter
defined by
$U\mbox{\rm diag}(0,\Delta E_{21},\Delta E_{31},\Delta E_{41})U^{-1}
+\mbox{\rm diag}(A_e,0,0,A_n)=
\tilde{U}{\rm diag}(\tilde{E}_1,\tilde{E}_2,\tilde{E}_3,\tilde{E}_4)
\tilde{U}^{-1}$ ($\Delta E_{jk}\equiv E_j-E_k\simeq
\Delta m^2_{jk}/2E\equiv (m^2_j-m^2_k)/2E$).
The matter potentials $A_e$, $A_n$ are given by $A_e=\sqrt{2}G_FN_e$,
$A_n=G_FN_n/\sqrt{2}$, where $N_e$ and $N_n$ are the density of
electrons and neutrinos, respectively.
The neutrino energy $E$ and
the baseline length $L$ which are typical
at a neutrino factory satisfy
$|\Delta m^2_{31}L/4E|\sim{\cal O}(1)$,
$|\Delta m^2_{21}L/4E|\ll 1$
and $|\Delta m^2_{41}L/4E|\gg 1$,
and the energy eigenvalues in this case
to the lowest order in the small mixing angles
and to first order in $|\Delta m^3_{31}|/|\Delta m^3_{41}|$ are
$\tilde{E}_1\sim \Delta E_{31}$, $\tilde{E}_2\sim 0$,
$\tilde{E}_3\sim A_e$, $\tilde{E}_4\sim \Delta E_{41}$.
It can be shown that the 4-th component $\tilde{X}^{\alpha\beta}_4$
in matter is the same as that in vacuum:
$\tilde{X}^{\alpha\beta}_4\simeq X^{\alpha\beta}_4$,
where the notation
$X^{\alpha\beta}_j\equiv U_{\alpha j}U^\ast_{\beta j}$
has been also introduced for the quantity in vacuum.
On the other hand, other three components are given by
\begin{eqnarray}
\tilde{X}^{\beta\alpha}_1
&=&
-\Delta \tilde{E}_{21}^{-1}\tilde{E}_{31}^{-1}
\{X^{\beta\alpha}_4\tilde{E}_2\tilde{E}_3
+(\tilde{E}_2+\tilde{E}_3)P^{\beta\alpha}
+Q^{\beta\alpha}\}
\nonumber\\
\tilde{X}^{\beta\alpha}_2
&=&
+\Delta \tilde{E}_{21}^{-1}\tilde{E}_{32}^{-1}
\{X^{\beta\alpha}_4\tilde{E}_3\tilde{E}_1
+(\tilde{E}_3+\tilde{E}_1)P^{\beta\alpha}
+Q^{\beta\alpha}\}
\nonumber\\
\tilde{X}^{\beta\alpha}_3
&=&
-\Delta \tilde{E}_{31}^{-1}\tilde{E}_{32}^{-1}
\{X^{\beta\alpha}_4\tilde{E}_1\tilde{E}_2
+(\tilde{E}_1+\tilde{E}_2)P^{\beta\alpha}
+Q^{\beta\alpha}\},
\label{x3}
\end{eqnarray}
where
\begin{eqnarray}
P^{\beta\alpha}&\equiv&
\{A(X^{ee}_4+X^{ss}_4/2)-{\cal A}_{\alpha\alpha}-{\cal A}_{\beta\beta}\}
X^{\beta\alpha}_4
+\Delta E_{31}X^{\beta\alpha}_3
+\Delta E_{21}X^{\beta\alpha}_2
\nonumber\\
Q^{\beta\alpha}&\equiv&
X^{\beta\alpha}_4
\{{\cal A}_{\alpha\alpha}^2+{\cal A}_{\alpha\alpha}{\cal A}_{\beta\beta}
+{\cal A}_{\beta\beta}^2
-A({\cal A}_{\alpha\alpha}+{\cal A}_{\beta\beta})(X^{ee}_4+X^{ss}_4/2)\}
\nonumber\\
&{\ }&
-\Delta E_{31}(\Delta E_{31}+{\cal A}_{\alpha\alpha}+{\cal A}_{\beta\beta})
X^{\beta\alpha}_3
\nonumber\\
&{\ }&
-\Delta E_{21}(\Delta E_{21}+{\cal A}_{\alpha\alpha}+{\cal A}_{\beta\beta})
X^{\beta\alpha}_2
\nonumber\\
&{\ }&
+A\Delta E_{31}(
X^{\beta e}_4X^{e\alpha}_3+X^{\beta e}_3X^{e\alpha}_4
+X^{\beta s}_4X^{s\alpha}_3+X^{\beta s}_3X^{s\alpha}_4)
\nonumber\\
&{\ }&
+A\Delta E_{21}(
X^{\beta e}_4X^{e\alpha}_2+X^{\beta e}_2X^{e\alpha}_4
+X^{\beta s}_4X^{s\alpha}_2+X^{\beta s}_2X^{s\alpha}_4).
\label{x}
\end{eqnarray}
In Eq. (\ref{x}) ${\cal A}_{\alpha\alpha}=
A_e\delta_{\alpha e}+A_n\delta_{\alpha s}$
is the matrix element of the matter potential,
and no sum is understood over the indices $\alpha, \beta$.
If the sterile neutrino mixings
$X^{\alpha\beta}_4~(\alpha=e, \mu, \tau)$
are small, then $\tilde{X}^{\beta\alpha}_j~(j=1,2,3)$
reproduce those for the standard three flavor case.
These sterile neutrino mixings $X^{\alpha\beta}_4$
appear in the coefficients $\tilde{X}^{\beta\alpha}_j~(j=1,2,3)$
in front of the sine factors $\sin^2({\Delta \tilde{E}_{jk}L / 2})$,
so we can get information
on the sterile neutrino mixings from
precise measurements of the coefficients
of the oscillation mode
$\sin^2(\Delta \tilde{E}_{jk}^2 L/4E)~(j,k=1,2,3)$,
which are the dominant contribution to the
probability.
We have evaluated sensitivity numerically by taking
matter effects into account, and the results are given in
Figs.\ref{aba:fig2}-\ref{aba:fig5}.
Since we have assumed $\Delta m_{41}^2>{\cal O}(0.1$eV$^2$),
the results for $\Delta m_{41}^2 < 0.1$ eV$^2$
are not given in the figures.
The advantage of measurements with the far detectors is that
sensitivity is independent of $\Delta m_{41}^2$
and it is good even for lower values of
$\Delta m_{41}^2$ in most cases.
In particular,
in the case of the golden channel
$\nu_e\rightarrow\nu_\mu$,
the far detectors improve the present
bound on $4|U_{e4}|^2|U_{\mu4}|^2$
by two orders of magnitude for
all the values of $\Delta m_{41}^2>{\cal O}(0.1$eV$^2$).
The neutrino factory with far detectors,
therefore, can provide a very powerful test
of the LSND anomaly.
Their disadvantage of measurements with the far detectors is that
sensitivity is not as good as
that of the near detectors, which will be
described in the next subsection, at the peak.

\subsection{Sensitivity of a neutrino factory with near detectors\cite{Donini:2001xy}}

In my talk I skipped the discussions on sensitivity
of a neutrino factory with near detectors, but
because of recent interest on the near detector issues,\cite{Tang:2009na,minsis,uamnsi}
I will describe sensitivity of measurements with
near detectors for the sake of completeness.
In Ref.~\refcite{Donini:2001xy} sensitivity of
a neutrino factory with near detectors
to sterile neutrino mixings was studied.
The setup used in this analysis is the following:
the muon energy is 20GeV,
the number of useful muons is
$2\times10^{20}$ $\mu^-$'s per year,
the measurements are supposed to continue
for 5 years, the volume of a magnetized iron
calorimeter at the distance $L$=40km is 40kton,
that of an emulsion cloud chamber at $L$=1km
is 1kton, and the statistical errors and
the backgrounds are taken into account.\footnote{
In this analysis the effects of the systematic
errors were not taken into account.
Their results can be refined 
in the future.}

At such short baselines,
$|\Delta m^2_{41}L/2E|\sim{\cal O}(1)\gg
|\Delta m^2_{31}L/2E|\gg|\Delta m^2_{21}L/2E|$
is satisfied, so the only relevant
mass squared difference is $\Delta m^2_{41}$.
So we have the
following oscillation probabilities:
\begin{eqnarray}
\hspace{-5mm}
P(\nu_e\rightarrow\nu_\mu)
&\simeq&
4\left|U_{e4}\right|^2\left|U_{\mu4}\right|^2
\,\sin^2  ( {\Delta m_{41}^2 L/4 E}  )
\nonumber\\
P(\nu_\mu\rightarrow\nu_\mu)
&\simeq&
1-4|U_{\mu4}|^2(1-|U_{\mu4}|^2)
\,\sin^2 ( {\Delta m_{41}^2 L/4 E}  )
\nonumber\\
P(\nu_e\rightarrow\nu_\tau)
&\simeq&
4\left|U_{e4}\right|^2\left|U_{\tau4}\right|^2
\,\sin^2  ( {\Delta m_{41}^2 L/4 E}  )
\nonumber\\
P(\nu_\mu\rightarrow\nu_\tau)
&\simeq&
4\left|U_{\mu4}\right|^2\left|U_{\tau4}\right|^2
\,\sin^2  ( {\Delta m_{41}^2 L/4 E}  )
\nonumber
\end{eqnarray}
Thus we can determine $4|U_{\alpha4}|^2|U_{\beta4}|^2$
or $4|U_{\mu4}|^2(1-|U_{\mu4}|^2)$
from the coefficient of the dominant oscillation
mode $\sin^2(\Delta m_{41}^2 L/4E)$.
The results are shown in Figs.\ref{aba:fig2}-\ref{aba:fig5}.  The
mass squared difference for which this neutrino factory setup has the
best performance depends on the baseline length $L$, and
in the present case it is approximately 10eV$^2$.
The advantage of measurements with the near detectors is that sensitivity
to the sterile neutrino mixings is very good at the peak
while their disadvantage is that sensitivity becomes
poorer for lower values of $\Delta m^2_{41}$.
From these results, we conclude that
the near and far detectors are complementary
in their performance.

\section{The CP phases due to new physics}
The results in the previous section suggest
that the discovery channel $\nu_\mu\to\nu_\tau$
may not be so powerful in giving the upper bound
on the mixing angles.
To see the role of the discovery channel,
let us now consider the effects of the CP phases
in neutrino oscillations.

\subsection{T violation in four neutrino schemes}
In matter T violation
$P_{\alpha\beta}-P_{\beta\alpha} \equiv
P(\nu_\alpha\to\nu_\beta)-P(\nu_\beta\to\nu_\alpha)$
is more useful than CP violation
$P(\nu_\alpha\to\nu_\beta)-P(\bar\nu_\alpha\to\bar\nu_\beta)$,
so I will discuss T violation in four neutrino schemes.\footnote{
Note that we are not claiming that T violation can be measured
experimentally for all the channels.
The oscillation probability can be always
decomposed into T conserving and T violating terms,
$P_{\alpha\beta}=(P_{\alpha\beta}+P_{\beta\alpha})/2
+(P_{\alpha\beta}-P_{\beta\alpha})/2$, and
the second term is proportional to $\sin\delta$
in the standard three flavor framework\cite{Naumov:1991ju}
in constant-density matter, as in the case in vacuum.
so T violation is phenomenologically suitable to examine $\delta$.
In the case of CP violation, on the other hand,
CP is violated in matter even if the CP phase vanishes.
In practice, people perform a numerical
analysis by fitting the hypothetical oscillation
probability to the full data including neutrinos
and anti-neutrinos, instead of measuring
$P(\nu_\alpha\to\nu_\beta)-P(\bar\nu_\alpha\to\bar\nu_\beta)$
or
$P(\nu_\alpha\to\nu_\beta)-P(\nu_\beta\to\nu_\alpha)$.
So discussions on CP violation or T violation
should be regarded as tools
to help us understand the results intuitively.}

In the three flavor scheme it is known that T violation is given by
\begin{eqnarray}
P_{\alpha\beta}-P_{\beta\alpha}
&=&-16\,\mbox{\rm Im}(\tilde{X}^{\beta\alpha}_1\tilde{X}^{\beta\alpha\ast}_2)\,
\sin\frac{\Delta \tilde{E}_{21}L}{2}
\sin\frac{\Delta \tilde{E}_{31}L}{2}
\sin\frac{\Delta \tilde{E}_{32}L}{2}.
\label{cp3}
\end{eqnarray}
The Jarlskog factor
Im$(\tilde{X}^{\beta\alpha}_1\tilde{X}^{\beta\alpha\ast}_2)$
can be written as\cite{Naumov:1991ju}
\begin{eqnarray}
\mbox{\rm Im}(\tilde{X}^{\beta\alpha}_1\tilde{X}^{\beta\alpha\ast}_2)
=\mbox{\rm Im}(X^{\beta\alpha}_1X^{\beta\alpha\ast}_2)
\Delta E_{21}\Delta E_{31}\Delta E_{32}
/\Delta \tilde{E}_{21}\Delta \tilde{E}_{31}\Delta \tilde{E}_{32}.
\label{j3}
\end{eqnarray}
If $|\Delta E_{31}L|\sim{\cal O}(1)$, then the
differences of the eigenvalues in this case are all
of ${\cal O}(\Delta E_{31})$
in the zeroth order in $\sin^2\theta_{13}$.
In that case the product of
the sine factors $\prod_{j<k} \sin(\Delta \tilde{E}_{j,k}L/2)$ is
of $\cal O$(1), but in Eq.(\ref{j3}) we have a suppression factor
$|\Delta E_{21}/\Delta\tilde{E}_{21}|\sim
|\Delta m^2_{21}/\Delta m^2_{31}|\sim 1/30$.

On the other hand, in the case of the four
neutrino schemes, Eq.(\ref{cp3}) is replaced by
\begin{eqnarray}
P_{\alpha\beta}-P_{\beta\alpha}
&=&-16\,\mbox{\rm Im}(\tilde{X}^{\beta\alpha}_1\tilde{X}^{\beta\alpha\ast}_2)\,
\sin\frac{\Delta \tilde{E}_{21}L}{2}
\sin\frac{\Delta \tilde{E}_{41}L}{2}
\sin\frac{\Delta \tilde{E}_{42}L}{2}
\nonumber\\
&{\ }&-16\,\mbox{\rm Im}(\tilde{X}^{\beta\alpha}_1\tilde{X}^{\beta\alpha\ast}_3)\,
\sin\frac{\Delta \tilde{E}_{31}L}{2}
\sin\frac{\Delta \tilde{E}_{41}L}{2}
\sin\frac{\Delta \tilde{E}_{43}L}{2}
\nonumber\\
&{\ }&-16\,\mbox{\rm Im}(\tilde{X}^{\beta\alpha}_2\tilde{X}^{\beta\alpha\ast}_3)\,
\sin\frac{\Delta \tilde{E}_{32}L}{2}
\sin\frac{\Delta \tilde{E}_{42}L}{2}
\sin\frac{\Delta \tilde{E}_{43}L}{2}
\label{cp4}
\end{eqnarray}
In the case of neutrino energy with
$|\Delta E_{31}L|\sim{\cal O}(1)$,
the dominant contribution in Eq.(\ref{cp4})
to the leading order in the small mixing angles
is given by
\begin{eqnarray}
P_{\alpha\beta}-P_{\beta\alpha}
\simeq \sum_{(j,k)=(1,2),(1,3),(2,3)}
4\,\mbox{\rm Im}(\tilde{X}^{\beta\alpha}_j\tilde{X}^{\beta\alpha\ast}_k)\,
\sin\Delta \tilde{E}_{jk}L,
\label{j4}
\end{eqnarray}
\fussy
where we have averaged over rapid oscillations
due to $\Delta m^2_{41}$, i.e.,
$\lim_{x\to\infty}\sin x\sin(x+\theta)$ = $\cos\theta/2$.
To compare T violation in the different schemes
or in different channels,
we have only to
compare the Jarlskog factors Im($X^{\beta\alpha}_jX^{\beta\alpha\ast}_k$)
in Eqs.(\ref{j4}) and (\ref{j3}),
since the sine factors $\prod_{j<k} \sin(\Delta \tilde{E}_{j,k}L/2)$
in Eq.(\ref{cp3}) and $\sin\Delta \tilde{E}_{jk}L$ in Eq.(\ref{j4}) are
both of ${\cal O}(1)$.
If the sterile neutrino mixing angles are roughly as large
as $\theta_{13}$, then it turns out the dominant contribution
to the Jarlskog factor comes from
Im$(X^{\beta\alpha}_3X^{\beta\alpha\ast}_4)$,
which appears
in Im$(\tilde{X}^{\beta\alpha}_j\tilde{X}^{\beta\alpha\ast}_k)$
with coefficients of ${\cal O}(1)$
if we plug Eq.(\ref{x3}) in Eq.(\ref{j4}).
Furthermore,
let us introduce a parametrization for the $4\times4$ mixing matrix
with three CP phases $\delta_\ell$:
\vglue -5mm
\begin{eqnarray}
    U =
    R_{34}(\theta_{34} ,\, 0) \; R_{24}(\theta_{24} ,\, 0) \;
    R_{23}(\theta_{23} ,\, \delta_3) \;
    R_{14}(\theta_{14} ,\, 0) \; R_{13}(\theta_{13} ,\, \delta_2) \; 
    R_{12}(\theta_{12} ,\, \delta_1) \,,
\nonumber
\end{eqnarray}
where $R_{ij}(\theta_{ij},\ \delta_l)$ are the $4\times4$ complex
rotation matrices defined by
\begin{eqnarray}
[R_{ij}(\theta_{ij},\ \delta_{l})]_{pq} = 
\left\{ 
\begin{array}{ll} \cos \theta_{ij} & p=q=i,j \\
1 & p=q \not= i,j \\
\sin \theta_{ij} \ e^{-i\delta_{l}} &	p=i;q=j \\
-\sin \theta_{ij} \ e^{i\delta_{l}} & p=j;q=i \\
0 & \mbox{\rm otherwise.}\end{array}
\right\}.
\nonumber
\end{eqnarray}
$\theta_{14}$ stands for the mixing angle
in short baseline reactor neutrino oscillations, and
$\theta_{24}$ ($\theta_{34}$) represents the ratio of the oscillation modes
due to $\Delta m^2_{31}$ and $\Delta m^2_{41}$
(the ratio of the
active and sterile neutrino oscillations) in the atmospheric neutrinos,
respectively.
In the limit when these extra mixing angles $\theta_{j4}~(j=1,2,3)$
become zero, $\delta_2$
becomes the standard CP phase in the three flavor scheme.
The explicit forms of the mixing matrix elements $U_{\alpha j}$
can be found in the Appendix A in Ref.~\refcite{Donini:2008wz}.

From the constraints of the short baseline
reactor experiments and the atmospheric neutrino data,
these angles are constrained as\cite{Donini:2007yf}
$\theta_{14}\lesssim 10^\circ$,
$\theta_{24}\lesssim 12^\circ$,
$\theta_{34}\lesssim 30^\circ$.
If we assume the upper bounds for
$\theta_{j4}~(j=1,2,3)$ and $\theta_{13}$,
for which we have
$\theta_{13}\lesssim 13^\circ$,
then together with the best fit values
for the solar and atmospheric oscillation angles
$\theta_{12}\simeq 30^\circ$,
$\theta_{23}\simeq 45^\circ$,
we obtain the following Jarlskog factor:
\begin{eqnarray}
4\left|\mbox{\rm Im}(X^{e\mu}_3X^{e\mu\ast}_4)
\right|_{\mbox{\scriptsize\rm 4flavor}}
&\simeq&
4|s_{23}s_{13}s_{14}s_{24}\sin(\delta_3-\delta_2)|
\lesssim 0.02\,|\sin(\delta_3-\delta_2)|
\nonumber\\
4\left|\mbox{\rm Im}(X^{\mu\tau}_3X^{\mu\tau\ast}_4)
\right|_{\mbox{\scriptsize\rm 4flavor}}
&\simeq&
4|s_{23}s_{24}s_{34}\sin\delta_3|\lesssim 0.2\,|\sin\delta_3|
\nonumber
\end{eqnarray}
for the (3+1)-scheme, where $c_{jk} \equiv \cos \theta_{jk}$ and
$s_{jk} \equiv \sin \theta_{jk}$.
These results should be compared with the
standard Jarlskog factor:
\begin{eqnarray}
4\left|\mbox{\rm Im}(X^{e\mu}_1X^{e\mu\ast}_2)
\right|_{\mbox{\scriptsize\rm 3flavor}}
&=&
(1/2)|c_{13}\sin2\theta_{12}\sin2\theta_{23}\sin2\theta_{13}\sin\delta|
\lesssim 0.2\,|\sin\delta|.
\nonumber
\end{eqnarray}
Notice that the Jarlskog factor is independent of
the flavor $(\alpha,\beta)$ in the three flavor case.
Assuming that all the CP phases are
maximal, i.e.,
$|\sin\delta_3|\sim |\sin(\delta_3-\delta_2)|\sim
|\sin\delta| \sim{\cal O}(1)$,
the ratio of T violation in
the (3+1)-scheme for the two channel
and that in the standard three flavor scheme is given by
\begin{eqnarray}
\left|P_{e\mu}-P_{\mu e}\right|_{\mbox{\scriptsize\rm 4flavor}}:
\left|P_{\mu\tau}-P_{\tau \mu}\right|_{\mbox{\scriptsize\rm 4flavor}}:
\left|P_{e\mu}-P_{\mu e}\right|_{\mbox{\scriptsize\rm 3flavor}}
\sim 0.02:0.2:0.006.
\nonumber
\end{eqnarray}
Note that the term which would reduce to the three flavor
T violation in the limit $\theta_{j4}\to0~(j=1,2,3)$
is contained in Eq.(\ref{j4}) as a subdominant contribution
which is suppressed by $|\Delta m^2_{21}/\Delta m^2_{31}|\sim 1/30$.
From this we see that dominant contribution to
T violation in the (3+1)-scheme
could be potentially much larger when measured with the discovery channel
than that in the (3+1)-scheme with the golden channel or
than that in the standard three flavor scheme.
In fact it was shown in Ref.~\refcite{Donini:2008wz} by a detailed analysis
that the CP phase may be measured
using the discovery and disappearance channels.

\subsection{CP violation in unitarity violation due to heavy fields}
In generic see-saw models the kinetic term gets modified after
integrating out the right handed neutrino and unitarity is expected to
be violated.\cite{Antusch:2006vwa}
In the case of the so-called minimal unitarity violation,
in which only three light neutrinos are involved and
sources of unitarity violation are assumed to appear only
in the neutrino sector,
deviation from unitarity is strongly constrained
from the rare decays of charged leptons.
Expressing the nonunitary mixing matrix $N$ as
$N=(1+\eta)U$, where $U$ is a unitary matrix while $\eta$
is a hermitian matrix which stands for deviation
from unitarity, the bounds are typically
$|\eta_{\alpha\beta}|<{\cal O}$(0.1\%)~\cite{Antusch:2008tz}.
The CP asymmetry in this scenario in the two flavor framework
can be expressed as\cite{FernandezMartinez:2007ms}
\begin{eqnarray}
\frac{P_{\alpha\beta}
-P_{\bar\alpha \bar\beta}}{P_{\alpha\beta}
+P_{\bar\alpha \bar\beta}}\sim 
\frac{ -4|\eta_{\alpha \beta}|\sin(\mbox{\rm arg}(\eta_{\alpha \beta}))}
{\sin(2\theta)\sin\left(\Delta EL/2\right)}\,.
\nonumber
\end{eqnarray}
The constraint on $\eta_{\mu \tau}$ is
weaker that than on $\eta_{e\mu}$, and
it was shown\cite{FernandezMartinez:2007ms,Antusch:2009pm} that
the CP violating phase arg$(\eta_{\alpha \beta})$ may be measured
at a neutrino factory with the discovery channel.

\section{Summary}
In this talk I described sensitivity of a neutrino
factory to the sterile neutrino mixings.
The golden channel $\nu_e\to\nu_\mu$ improves
the present upper bound on
$4|U_{e4}|^2|U_{\mu4}|^2$ by two orders of
magnitude, and provides a powerful test
for the LSND anomaly.
It is emphasized that $\tau$ detection
at a neutrino factory is important to check unitarity,
and the discovery channel $\nu_\mu\to\nu_\tau$
is one of the promising channels to look for
physics beyond the standard three flavor scenario.
We may be able measure the new CP violating phase
using this channel in the sterile neutrino
schemes and in the scenario with unitarity violation
due to heavy particles.

\section*{Acknowledgments}
I would like to thank H.V. Klapdor-Kleingrothaus,
R.D. Viollier and other organizers for invitation and hospitality
during the conference.  I would also like to thank A.~Donini, K.~Fuki,
J.~Lopez-Pavon and D.~Meloni for collaboration on Ref.~\refcite{Donini:2008wz}.
This research was
supported in part by a Grant-in-Aid for Scientific Research of the
Ministry of Education, Science and Culture, \#21540274.


\begin{thebibliography}{1}

\bibitem{Amsler:2008zz}
  C.~Amsler {\it et al.}  [Particle Data Group],
  Phys.\ Lett.\  B {\bf 667} (2008) 1.

\bibitem{Fogli:2008jx}
  G.~L.~Fogli, E.~Lisi, A.~Marrone, A.~Palazzo and A.~M.~Rotunno,
  Phys.\ Rev.\ Lett.\  {\bf 101} (2008) 141801
  [arXiv:0806.2649 [hep-ph]].

\bibitem{Fogli:2008cx}
  G.~L.~Fogli, E.~Lisi, A.~Marrone, A.~Palazzo and A.~M.~Rotunno,
  arXiv:0809.2936 [hep-ph].

\bibitem{Ge:2008sj}
  H.~L.~Ge, C.~Giunti and Q.~Y.~Liu,
  arXiv:0810.5443 [hep-ph].

\bibitem{Fogli:2009ce}
  G.~L.~Fogli, E.~Lisi, A.~Marrone, A.~Palazzo and A.~M.~Rotunno,
  arXiv:0905.3549 [hep-ph].

\bibitem{GonzalezGarcia:2010er}
  M.~C.~Gonzalez-Garcia, M.~Maltoni and J.~Salvado,
  arXiv:1001.4524 [hep-ph].

\bibitem{Bandyopadhyay:2007kx}
  A.~Bandyopadhyay {\it et al.}  [ISS Physics Working Group],
  Rept.\ Prog.\ Phys.\  {\bf 72}, 106201 (2009)
  [arXiv:0710.4947 [hep-ph]].

\bibitem{Itow:2001ee}
  Y.~Itow {\it et al.}  [The T2K Collaboration],
  arXiv:hep-ex/0106019.

\bibitem{Ishitsuka:2005qi}
  M.~Ishitsuka, T.~Kajita, H.~Minakata and H.~Nunokawa,
  Phys.\ Rev.\  D {\bf 72}, 033003 (2005)
  [arXiv:hep-ph/0504026].

\bibitem{Hagiwara:2005pe}
  K.~Hagiwara, N.~Okamura and K.~i.~Senda,
  Phys.\ Lett.\  B {\bf 637}, 266 (2006)
  [Erratum-ibid.\  B {\bf 641}, 486 (2006)]
  [arXiv:hep-ph/0504061].

\bibitem{Ayres:2004js}
  D.~S.~Ayres {\it et al.}  [NOvA Collaboration],
  arXiv:hep-ex/0503053.

\bibitem{Maricic:2010zza}
  J.~Maricic  [LBNE DUSEL Collaboration],
  J.\ Phys.\ Conf.\ Ser.\  {\bf 203}, 012109 (2010).

\bibitem{Zucchelli:2002sa}
  P.~Zucchelli,
  Phys.\ Lett.\  B {\bf 532} (2002) 166.

\bibitem{Geer:1997iz}
  S.~Geer,
  Phys.\ Rev.\  D {\bf 57} (1998) 6989
  [Erratum-ibid.\  D {\bf 59} (1999) 039903]
  [arXiv:hep-ph/9712290].

\bibitem{giunti}
Some of the references on sterile neutrinos are found at the 
Neutrino Unbound web page, by C. Giunti and M. Laveder,
{\tt http://www.nu.to.infn.it/Sterile\_Neutrinos/}.

\bibitem{belle:0000}
Belle experiment,
{\tt http://belle.kek.jp/}.

\bibitem{babar:0000}
Babar experiment,
{\tt http://www-public.slac.stanford.edu/babar/}.

\bibitem{Wolfenstein:1977ue}
  L.~Wolfenstein,
  Phys.\ Rev.\  D {\bf 17}, 2369 (1978).

\bibitem{Guzzo:1991hi}
  M.~M.~Guzzo, A.~Masiero and S.~T.~Petcov,
  Phys.\ Lett.\  B {\bf 260} (1991) 154.

\bibitem{Roulet:1991sm}
  E.~Roulet,
  Phys.\ Rev.\  D {\bf 44} (1991) 935.

\bibitem{Grossman:1995wx}
  Y.~Grossman,
  Phys.\ Lett.\  B {\bf 359} (1995) 141
  [arXiv:hep-ph/9507344].

\bibitem{Antusch:2006vwa}
  S.~Antusch, C.~Biggio, E.~Fernandez-Martinez, M.~B.~Gavela and J.~Lopez-Pavon,
  JHEP {\bf 0610} (2006) 084
  [arXiv:hep-ph/0607020].

\bibitem{Antusch:2008tz}
  S.~Antusch, J.~P.~Baumann and E.~Fernandez-Martinez,
  Nucl.\ Phys.\  B {\bf 810}, 369 (2009)
  [arXiv:0807.1003 [hep-ph]].

\bibitem{Donini:2007yf}
  A.~Donini, M.~Maltoni, D.~Meloni, P.~Migliozzi and F.~Terranova,
  JHEP {\bf 0712}, 013 (2007)
  [arXiv:0704.0388 [hep-ph]].

\bibitem{Athanassopoulos:1996jb}
  C.~Athanassopoulos {\it et al.}  [LSND Collaboration],
  Phys.\ Rev.\ Lett.\  {\bf 77} (1996) 3082
  [arXiv:nucl-ex/9605003].

\bibitem{Athanassopoulos:1997pv}
  C.~Athanassopoulos {\it et al.}  [LSND Collaboration],
  Phys.\ Rev.\ Lett.\  {\bf 81} (1998) 1774
  [arXiv:nucl-ex/9709006].

\bibitem{Aguilar:2001ty}
  A.~Aguilar {\it et al.}  [LSND Collaboration],
  Phys.\ Rev.\  D {\bf 64} (2001) 112007
  [arXiv:hep-ex/0104049].

\bibitem{AguilarArevalo:2007it}
  A.~A.~Aguilar-Arevalo {\it et al.}  [The MiniBooNE Collaboration],
  Phys.\ Rev.\ Lett.\  {\bf 98}, 231801 (2007)
  [arXiv:0704.1500 [hep-ex]].

\bibitem{Smirnov:2006bu}
  A.~Y.~Smirnov and R.~Zukanovich Funchal,
  Phys.\ Rev.\  D {\bf 74}, 013001 (2006)
  [arXiv:hep-ph/0603009].

\bibitem{Kim:1978xk}
  J.~E.~Kim,
  Phys.\ Rev.\ Lett.\  {\bf 41}, 360 (1978).

\bibitem{Sorel:2003hf}
  M.~Sorel, J.~M.~Conrad and M.~Shaevitz,
  Phys.\ Rev.\  D {\bf 70}, 073004 (2004)
  [arXiv:hep-ph/0305255].

\bibitem{Maltoni:2007zf}
  M.~Maltoni and T.~Schwetz,
  Phys.\ Rev.\  D {\bf 76}, 093005 (2007)
  [arXiv:0705.0107 [hep-ph]].

\bibitem{Dydak:1983zq}
  F.~Dydak {\it et al.},
  Phys.\ Lett.\  B {\bf 134}, 281 (1984).

\bibitem{Declais:1994su}
  Y.~Declais {\it et al.},
  Nucl.\ Phys.\  B {\bf 434}, 503 (1995).

\bibitem{Okada:1996kw}
  N.~Okada and O.~Yasuda,
  Int.\ J.\ Mod.\ Phys.\  A {\bf 12}, 3669 (1997)
  [arXiv:hep-ph/9606411].

\bibitem{Bilenky:1996rw}
  S.~M.~Bilenky, C.~Giunti and W.~Grimus,
  Eur.\ Phys.\ J.\  C {\bf 1}, 247 (1998)
  [arXiv:hep-ph/9607372].

\bibitem{Karagiorgi:2009nb}
  G.~Karagiorgi, Z.~Djurcic, J.~M.~Conrad, M.~H.~Shaevitz and M.~Sorel,
  Phys.\ Rev.\  D {\bf 80}, 073001 (2009)
  [Erratum-ibid.\  D {\bf 81}, 039902 (2010)]
  [arXiv:0906.1997 [hep-ph]].

\bibitem{Maltoni:2004ei}
  M.~Maltoni, T.~Schwetz, M.~A.~Tortola and J.~W.~F.~Valle,
  New J.\ Phys.\  {\bf 6}, 122 (2004)
  [arXiv:hep-ph/0405172v6].

\bibitem{Donini:2008wz}
  A.~Donini, K.~i.~Fuki, J.~Lopez-Pavon, D.~Meloni and O.~Yasuda,
  JHEP {\bf 0908}, 041 (2009)
  [arXiv:0812.3703 [hep-ph]].

\bibitem{Donini:2001xy}
  A.~Donini and D.~Meloni,
  Eur.\ Phys.\ J.\  C {\bf 22}, 179 (2001)
  [arXiv:hep-ph/0105089].

\bibitem{Tang:2009na}
  J.~Tang and W.~Winter,
  Phys.\ Rev.\  D {\bf 80}, 053001 (2009)
  [arXiv:0903.3039 [hep-ph]].

\bibitem{minsis}
\fussy
Main Injector Non Standard Interactions Search,\\
{\tt http://www-off-axis.fnal.gov/MINSIS/}.

\bibitem{uamnsi}
\fussy
Madrid Neutrino NSI Workshop,
UAM, Madrid, 10-11 December 2009,\\
{\tt http://www.ft.uam.es/workshops/neutrino/default.html}.

\bibitem{BurguetCastell:2001ez}
  J.~Burguet-Castell, M.~B.~Gavela, J.~J.~Gomez-Cadenas, P.~Hernandez and O.~Mena,
  Nucl.\ Phys.\  B {\bf 608}, 301 (2001)
  [arXiv:hep-ph/0103258].

\bibitem{Minakata:2001qm}
  H.~Minakata and H.~Nunokawa,
  JHEP {\bf 0110}, 001 (2001)
  [arXiv:hep-ph/0108085].

\bibitem{Fogli:1996pv}
  G.~L.~Fogli and E.~Lisi,
  Phys.\ Rev.\ D {\bf 54} (1996) 3667 [arXiv:hep-ph/9604415];

\bibitem{Barger:2001yr}
  V.~Barger, D.~Marfatia and K.~Whisnant,
  Phys.\ Rev.\  D {\bf 65}, 073023 (2002)
  [arXiv:hep-ph/0112119].

\bibitem{donini-ids4}
A.~Donini, talk at the 0th IDS plenary Meeting, CERN 29-31 March 2007,\\
{\tt http://www.hep.ph.ic.ac.uk/ids/communication/cern-2007-03-29/slides/\\
IDStalk-Donini.pdf}.

\bibitem{CerveraVillanueva:2008zz}
  A.~Cervera-Villanueva,
  AIP Conf.\ Proc.\  {\bf 981}, 178 (2008).

\bibitem{Donini:2002rm}
  A.~Donini, D.~Meloni and P.~Migliozzi,
  Nucl.\ Phys.\  B {\bf 646} (2002) 321
  [arXiv:hep-ph/0206034].
  
\bibitem{Autiero:2003fu}
  D.~Autiero {\it et al.},
  Eur.\ Phys.\ J.\  C {\bf 33} (2004) 243
  [arXiv:hep-ph/0305185].

\bibitem{Abe:2007bi}
  T.~Abe {\it et al.}  [ISS Detector Working Group],
  arXiv:0712.4129 [physics.ins-det].

\bibitem{Kimura:2002hb}
K.~Kimura, A.~Takamura and H.~Yokomakura,
Phys.\ Lett.\ B {\bf 537}, 86 (2002)
[arXiv:hep-ph/0203099].

\bibitem{Kimura:2002wd}
K.~Kimura, A.~Takamura and H.~Yokomakura,
Phys.\ Rev.\ D {\bf 66}, 073005 (2002)
[arXiv:hep-ph/0205295].

\bibitem{Xing:2005gk}
  Z.~z.~Xing and H.~Zhang,
  Phys.\ Lett.\  B {\bf 618} (2005) 131
  [arXiv:hep-ph/0503118].

\bibitem{Yasuda:2007jp}
  O.~Yasuda,
  arXiv:0704.1531 [hep-ph].
  
\bibitem{Zhang:2006yq}
  H.~Zhang,
  Mod.\ Phys.\ Lett.\  A {\bf 22}, 1341 (2007)
  [arXiv:hep-ph/0606040].

\bibitem{Dighe:2007uf}
  A.~Dighe and S.~Ray,
  Phys.\ Rev.\  D {\bf 76}, 113001 (2007)
  [arXiv:0709.0383 [hep-ph]].

\bibitem{Astier:2001yj}
  P.~Astier {\it et al.}  [NOMAD Collaboration],
  Nucl.\ Phys.\  B {\bf 611}, 3 (2001)
  [arXiv:hep-ex/0106102].

\bibitem{Eskut:2007rn}
  E.~Eskut {\it et al.}  [CHORUS Collaboration],
  Nucl.\ Phys.\  B {\bf 793}, 326 (2008)
  [arXiv:0710.3361 [hep-ex]].

\bibitem{Stockdale:1984cg}
  I.~E.~Stockdale {\it et al.},
  Phys.\ Rev.\ Lett.\  {\bf 52}, 1384 (1984).

\bibitem{Naumov:1991ju}
  V.~A.~Naumov,
  Int.\ J.\ Mod.\ Phys.\  D {\bf 1}, 379 (1992).

\bibitem{FernandezMartinez:2007ms}
  E.~Fernandez-Martinez, M.~B.~Gavela, J.~Lopez-Pavon and O.~Yasuda,
  Phys.\ Lett.\  B {\bf 649}, 427 (2007)
  [arXiv:hep-ph/0703098].

\bibitem{Antusch:2009pm}
  S.~Antusch, M.~Blennow, E.~Fernandez-Martinez and J.~Lopez-Pavon,
  Phys.\ Rev.\  D {\bf 80}, 033002 (2009)
  [arXiv:0903.3986 [hep-ph]].

\end{thebibliography}
\end{document}